\documentclass[3p,times,twocolumn]{elsarticle}

\usepackage{ecrc}

\volume{00}

\firstpage{1}

\journalname{Nuclear Physics B Proceedings Supplement}

\runauth{}

\jid{nuphbp}

\jnltitlelogo{Nuclear Physics B Proceedings Supplement}

\usepackage{amssymb}

\usepackage[figuresright]{rotating}

\begin{document}

\begin{frontmatter}



\dochead{}

\title{Characterization of the front-end EASIROC for read-out of SiPM in the ASTRI camera}

 \author[label1]{D. Impiombato}
\ead{Domenico.Impiombato@ifc.inaf.it}
\author[label1]{S. Giarrusso}
\author[label2]{M. Belluso}
\author[label2]{S. Bilotta}
\author[label2]{G. Bonanno}
\author[label1]{O. Catalano}
\author[label2]{A. Grillo}
\author[label1]{G. La Rosa}
\author[label2]{D. Marano}
\author[label1]{T. Mineo}
\author[label1]{F. Russo}
\author[label1] {G. Sottile}
\author{on behalf of the ASTRI Collaboration}

 \address[label1]{INAF, Istituto di Astrofisica Spaziale e Fisica Cosmica di Palermo, via U. La Malfa 153,
I-90146 Palermo, Italy}
 \address[label2]{INAF, Osservatorio Astrofisico di Catania, via S. Sofia 78, I-95123 Catania, Italy}

\begin{abstract}
The design and realization of a prototype for the Small-Size class Telescopes of the Cherenkov Telescope Array 
is one of the cornerstones of the ASTRI project.  The prototype will adopt a focal plane camera based on 
Silicon Photo-Multiplier sensors that coupled with a dual mirror optics configuration represents an innovative 
solution for the detection of Atmospheric Cherenkov light.  These detectors can be read by the Extended Analogue 
Silicon Photo-Multiplier Integrated Read Out Chip (EASIROC) equipped with 32-channels. In this paper, 
we report some preliminary results on measurements aimed to evaluate EASIROC 
capability of autotriggering and measurements of the trigger time walk, jitter, DAC linearity and trigger
efficiency vs the injected charge. Moreover, the dynamic  range of the ASIC is also reported. 
\end{abstract}


\end{frontmatter}


\section{Introduction}

\begin{figure}
\centering
\includegraphics[width=7.1cm]{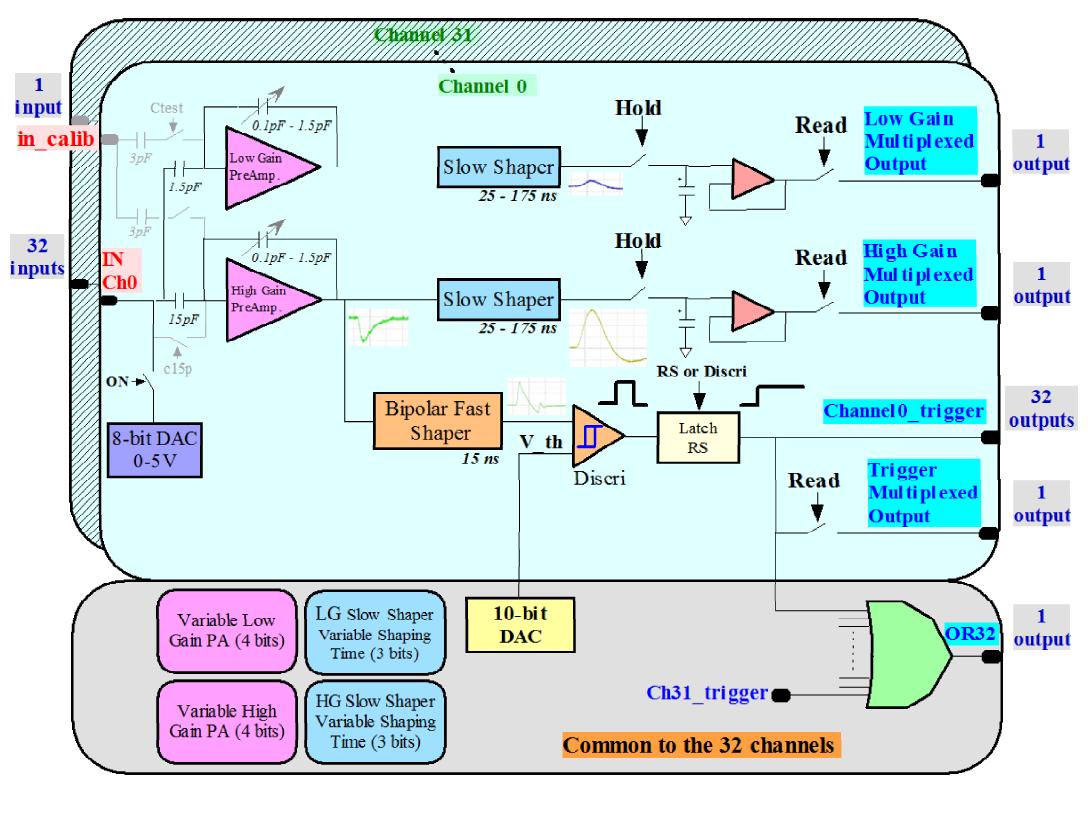}
\caption{Architecture of the EASIROC front-end (Omega courtesy)}
\label{fig1}
\end{figure}

ASTRI (Astrofisica con Specchi a Tecnologia Replicante Italiana) \cite{canestrari11} is a Flagship
Project financed by the Italian Ministry of Education,
University and Research (MIUR) and led by the Italian National Institute of Astrophysics (INAF).
 Main object of the project is the design and implementation of an end-to-end Small Size Telescope prototype 
 for the Cherenkov Telescope Array (CTA) \cite{cta10}.
The ASTRI prototype will adopt a wide field optical system in a Schwarzschild-Couder
configuration to explore the VHE range (1-100 TeV) of the electromagnetic spectrum and
a camera at the focal plane based on Silicon Photo-Multiplier (SiPM) sensors. This configuration  is  an
innovative solution for the detection of Cherenkov light that requires
high sensitivity in the 300--700 nm band and fast temporal response. SiPMs can be read
by the Extended Analogue Silicon Photo-Multiplier Integrated Read Out Chip (EASIROC) \cite{callier11}
equipped with 32-channels each with the capability of measuring charge from 1 to 2000
photoelectrons with a SiPM gain of 10$^{6}$ that is the gain we adopted in our measurements.
 A schematic of EASIROC is shown in Fig.~\ref{fig1}.

An Evaluation Board has been designed and built by Omega Micro\footnote{http://omega.in2p3.fr}
to test the functional characteristics and performance of this  Application Specific Integrated Circuit (ASIC).
The gain of each channel can be individually adjusted by varying the bias voltage, in a range of 
4.5V, with an 8 bit DAC(Digital to Analog Converter).
The SiPM signal is splitted and conveyed in two separate chains, high and low gain, in order to measure charges from
160fC up to 320pC allowing a dynamical range of 2000. 
Each of the two chains is composed by an adjustable gain preamplifier followed by a tunable 
shaper and a track and hold circuit.
A third chain is implemented to generate a trigger using a fast shaper followed by a discriminator, 
whose threshold is set by a 10-bit DAC, common to all 32 channels.
The power consumption is lower than 5mW/channel and unused features can be set $OFF$ to furtherly
save power.

In this paper,  we report preliminary results on the  measurements of the EASIROC output analogue 
signals to evaluate the capability of autotriggering.  
We present on the trigger time walk and jitter of the output signal, on the DAC linearity and on the
trigger efficiency vs the injected charge.
The paper presents also results on the the dynamic range of the ASIC as function of the injected charges.
\label{}



 \section{Trigger}

The trigger chain is derived from the High Gain preamplifier and is composed by a
dedicated 15ns fast shaper followed by a discriminator which provides the trigger signal(see Fig.~\ref{fig1}).
The threshold of the discriminators is common to all 32 channels and is set by a 10-bit DAC.
The DAC output range goes from 1.06V up to 2.38V, in steps of 1.28mV. 

\begin{figure}[!htbp]
\centering
\includegraphics[width=7.1cm]{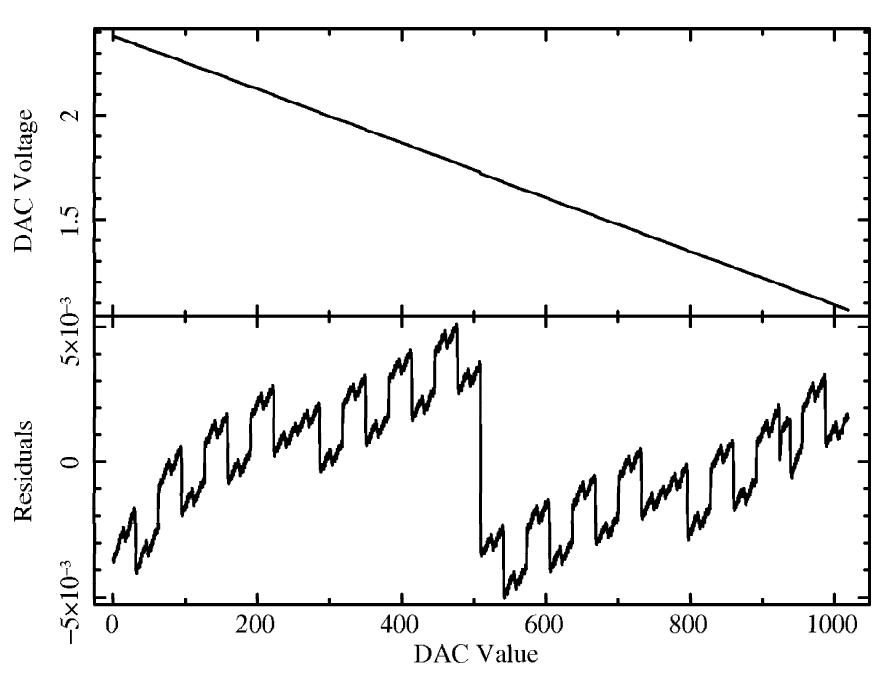}
\caption{\label{fig2}10-bit Input DAC Linearity}
\end{figure}

Figure~\ref{fig2} shows in the top panel the evolution of the DAC Voltage output
as a function of the input threshold and in the bottom panel the residuals of the fit
with a line. Residuals are of the order of 0.5$\%$.

\begin{figure}[!htbp]
\centering
\includegraphics[width=7.1cm]{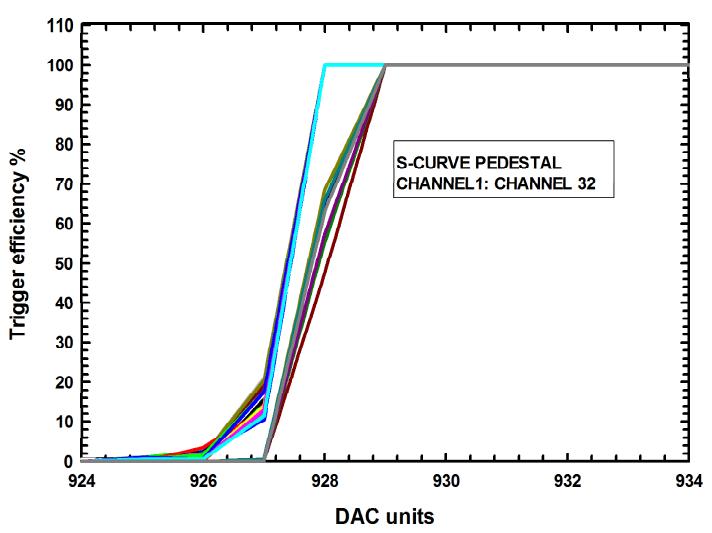}
\caption{\label{fig3}Pedestal S-curves for channel 1 to 32}
\end{figure}

\begin{figure}[!htbp]
\centering
\includegraphics[width=8cm]{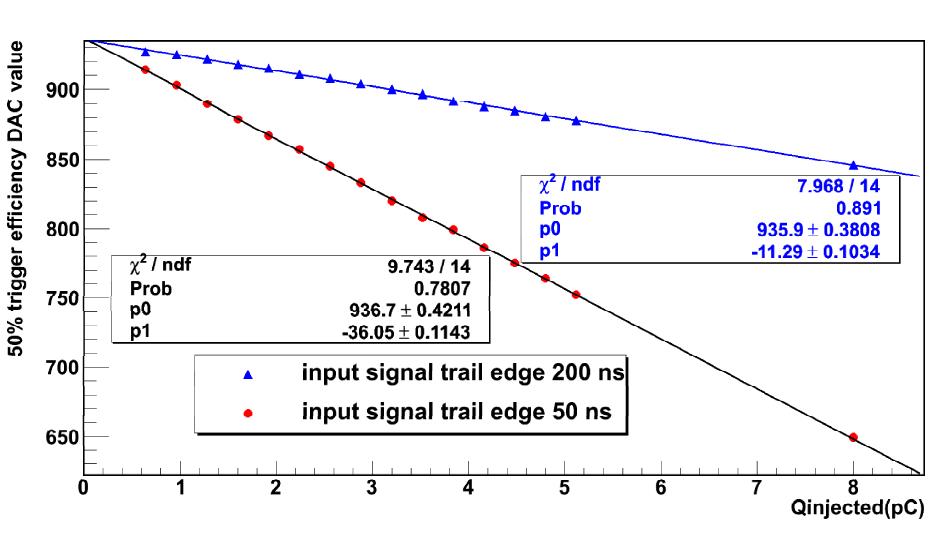}
\caption{50\% trigger efficiency threshold versus injected charge from 1 to 8 pC
 for a fixed channel}
\label{fig4}
\end{figure}

In order to evaluate the pedestal level of the trigger discriminators, the threshold level 
for all 32 channels was scanned without input signal.
As presented in the Fig.~\ref{fig3}, the resulted curve, called S-curve, shows that the pedestal level
measured at the 50$\%$ of trigger efficiency,
is the same for all the channels with a spread of about one DAC count.
The trigger efficiency was then evaluated scanning the threshold, in one channel,
by changing the injected charge and keeping constant the preamplifier gain at the same shaping time.
The measurements were carried out using two input SiPM-like signals  both with triangular shapes
with 1 ns of rise time.  The  decay time of the signal
was set to 50ns  and 200 ns, respectively.
We plotted the DAC channels corresponding to 50$\%$ efficiency, 
evaluated from the S-curves with an error of one channel, as
function of the injected charge. 
The linearity of the trigger (see Fig.~\ref{fig4}), is satisfactory and in line with the front-end electronics requirements. 

The trigger walk time and jitter were investigated injecting pulses  and
measuring at the oscilloscope the delay between the chip trigger output and the strobe
of the generator. Two sets of measurements were performed injecting 
charge through the calibration channel ({\itshape in\_calib}) with ASIC gain of 30. 
In this preliminary measurements the injected charge ranges from 0.48 up to 2.72pC (3pe up to 17pe) 
and from 1.6 up to 4.48pC (10pe up to 28pe) for trail edge input signals of 50 ns and 200 ns, 
respectively.

\begin{figure}[!h]
\centering
\includegraphics[angle=0,width=7.1cm]{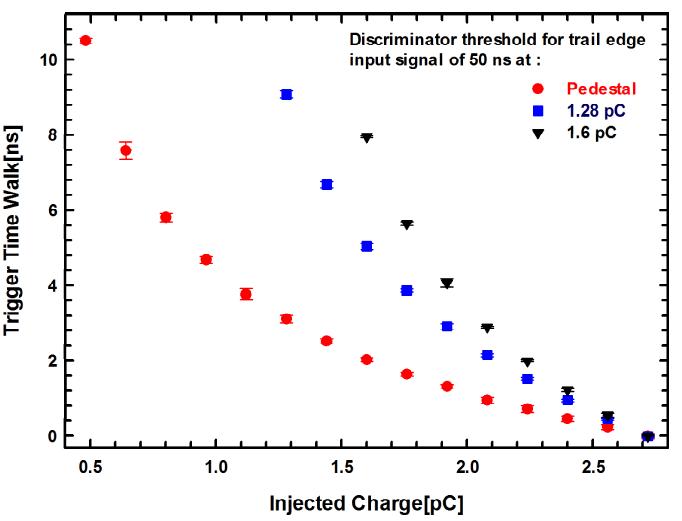} 
\includegraphics[angle=0,width=7.1cm]{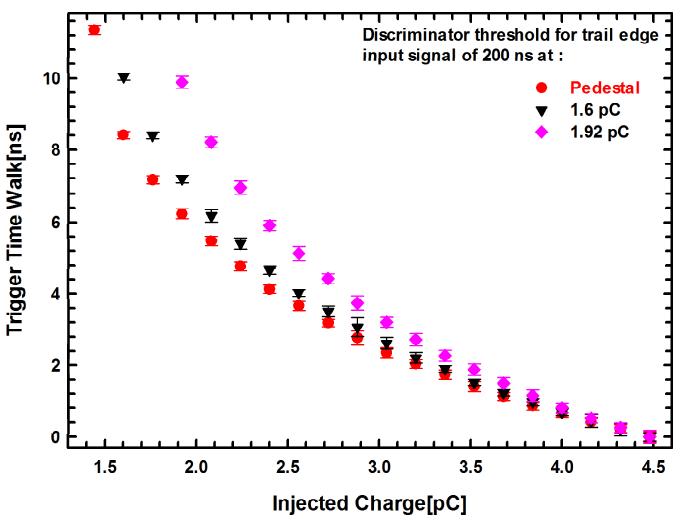} 
\caption{Trigger time walk for a threshold value above the pedestal as a function of the injected charge and for two input signals, 50ns (top panel) and 200ns (bottom panel), respectively.}
\label{fig5}
\end{figure}

\begin{figure}[!h]
\centering
\includegraphics[width=7.1cm]{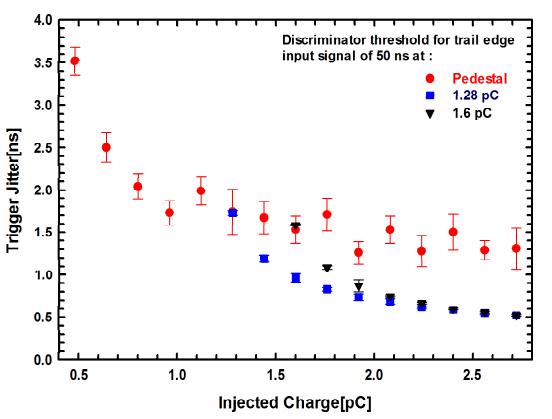}
\includegraphics[width=7.1cm]{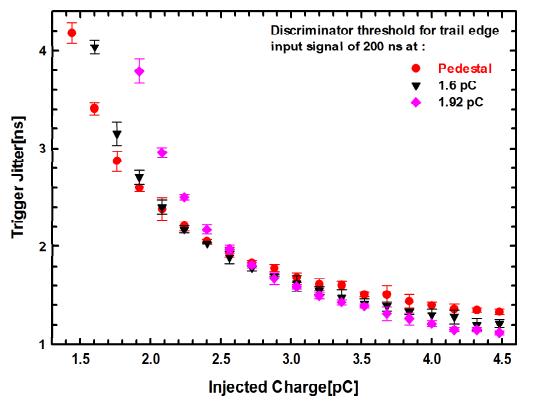} 
\caption{Trigger time jitter for a threshold value above the pedestal as a function of the injected charge and for two input signals, 50ns (top panel) and 200ns (bottom panel), respectively.}
\label{fig6}
\end{figure}

The results of the time walk and jitter measurements are presented in Fig.~\ref{fig5} and \ref{fig6}
for values of the injected 
As preliminary result, we find that the jitter is always lower than $\sim$4ns and the maximum walk time is $\sim$10ns.

\newpage
\section{Signal slow shaper and Charge linearity}
\begin{figure}[!t]
\centering
\includegraphics[width=7.1cm]{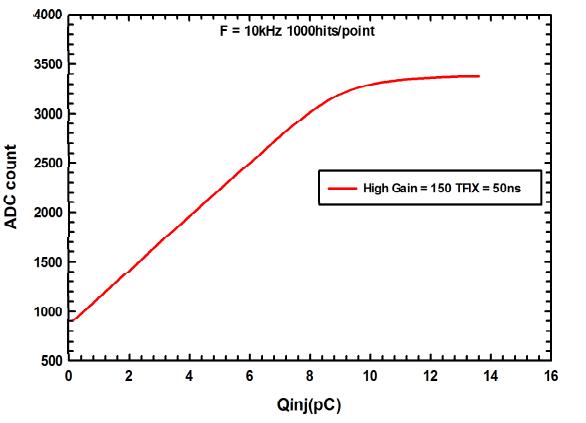}
\caption{Charge measurement up to 85 pe}
\label{fig7}
\end{figure}

\begin{figure}[!t]
\centering
\includegraphics[width=7.1cm]{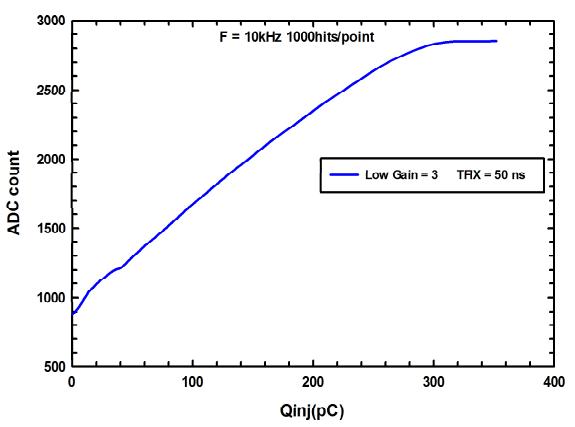}
\caption{Charge measurement up to 2200 pe}
\label{fig8}
\end{figure}

As shown in Fig.~\ref{fig1}, the signal coming from the detector is routed to a CRRC$^{2}$ slow shaper that provides a
charge measurement for both High Gain (1-85 pe) and Low Gain (1-2200 pe). 
The shaping time is fixed to 50 ns. The signal at the output of EASIROC is sampled by an external
Anologue to Digital Converter (ADC).
The linearity of the system was checked plotting the maximum of the peak versus the charge injected.
Results are shown in Figure \ref{fig7} and \ref{fig8}.  The linearity of the High Gain chain is 
maintained up to $\sim$ 8 pC, while the Low Gain shows a non linear bump for charge below 50pC.

\section{Conclusion}
This first preliminary set of measurements showed that EASIROC performance closely matches the 
ASTRI focal plane requirements and can be used for the telescope prototype. 
The problems on the linearity are not present in an update version of 
the chip, kindly provided us by OMEGA Micro.
However, to adopt this chip for CTA,  minor modifications of the ASIC design 
should be adopted.

 \section*{Acknowledgements}

We are deeply grateful to S. Callier, C. De La Taille, and L. Raux of the
OMEGA group at Orsay, for useful discussions and suggestions.

\nocite{*}
\bibliographystyle{elsarticle-num}
\bibliography{martin}








\end{document}